# The structure and dynamics of solvated electron in polar liquids. ¶

Ilya A. Shkrob [1)]

*Chemistry Division , Argonne National Laboratory, Argonne, IL 60439*

¶ Work performed under the auspices of the Office of Basic Energy Sciences, Division of Chemical Science, US-DOE under contract number DE-AC-02-06CH11357.

(1) e-mail: shkrob@anl.gov, tel: 630-2529516

## 1. INTRODUCTION

In this chapter, the recent progress in understanding of the nature and dynamics of excess (solvated) electrons in molecular fluids composed of polar molecules with no electron affinity (EA), such as liquid water (hydrated electron, $e^-_{hyd}$ ) and aliphatic alcohols, is examined. We have recently reviewed the literature on solvated electron in liquefied ammonia [1] and saturated hydrocarbons [2] and we address the reader to these publications for a brief introduction to the excess electron states in such liquids. We narrowed this review to bulk neat liquids and (to a much lesser degree) large water anion clusters in the gas phase that are useful reference systems for solvated electrons in the bulk. The excess electrons trapped by supramolecular structures (including single macrocycle molecules [3,4]), such as clusters of polar molecules [5,6] and water pools of reverse micelles [7,8] in nonpolar liquids and complexes of the electrons with cations [9] in concentrated salt solutions, are examined elsewhere.

This discourse echoes the themes addressed in our recent review on the properties of uncommon solvent anions. [10] We do not pretend to be comprehensive or inclusive, as the literature on electron solvation is vast and rapidly increasing. This increase is currently driven by ultrafast laser spectroscopy studies of electron injection and relaxation dynamics, and by gas phase studies of anion clusters by photoelectron and IR spectroscopy. Despite the great importance of the solvated/hydrated electron for radiation chemistry (as this species is a common reducing agent in radiolysis of liquids and solids), pulse radiolysis studies of solvated electrons are becoming less frequent perhaps due to the insufficient time resolution of the method (picoseconds) as compared to state-of-the-art laser studies (time resolution to 5 fs [11]). The welcome exception are the recent spectroscopic and kinetic studies of hydrated electrons in supercritical [12,13] and supercooled water. [14] As the theoretical models [12] for high-temperature hydrated electrons and the reaction mechanisms for these species are still debated, we will exclude such extreme conditions from this review.

Over the last 15 years, there was rapid progress in understanding the properties of solvated/hydrated electron. The advances were made simultaneously in many areas. First, it became possible to study the energetics and IR spectra of relatively large ammonia and water (and now methanol [15]) anion clusters in the gas phase. Very recently, pump-probe studies of such clusters have begun. [16,17] Second, resonance Raman spectra of solvated electrons in water and alcohols were obtained; these spectra provide direct insight into their structure. [18-22] Only magnetic resonance (electron paramagnetic resonance (EPR) and electron spin echo electron modulation (ESEEM)) studies carried out in the 1970s and the 1980s have provided comparably detailed picture of electron localization. [23] Third, there were multiple studies of ultrafast electron localization in water, alcohols, ethers, and ionic liquids following either photo- [24-28] or radiolytic [29] ionization of the solvent or electron detachment from a solvated anion (e.g., [30]) Numerous pump-probe studies of $e^-_{hyd}$ [11,31-35] and solvated electron in alcohols [36] and ethers, such as tetrahydrofuran (THF), [37] have been carried out. These studies addressed the initial stages of electron localization that are still not fully understood. Forth, theoretical (dynamic) models of electron solvation rapidly grew in sophistication and realism. In addition to one-electron models, both adiabatic and nonadiabatic, such as path integral, [38-40] mixed quantum-classical (MQC) molecular dynamics (MD), [41-43] mobile basis sets, [44-46] *ab initio* and density functional (DFT) approaches have been developed, chiefly for small and medium size water anion clusters. [47-49] Recently, these many-electron approaches began to be applied to solvated electrons in bulk liquids, using Car-Parrinello (CP) [50] or Born-Oppenheimer MD and hybrid MQC/MD:DFT calculations. [51] Meanwhile, the one-electron MQC MD methods are advancing towards more accurate representation of nonadiabatic transitions, decoherence effects [52] and quantum effects involving solvent degrees of freedom. [53] Fifth, these methods are presently applied to solvents of medium and low polarity, such as THF [54] and the alkanes, [2,4] where the electron dynamics and energetics are qualitatively different from those in water, and to dielectrons in water. [55]

While the dynamic studies of electron solvation are very important for understanding its chemistry, interpreting these dynamics is conditional on having the accurate picture of the ground state of the solvated electron. The point that is made in this review is that the current paradigm of the solvated electron as "a particle in a box" that informed the studies of the solvated electron for over 60 years [56-60] needs to be reassessed, despite its many successes. In place of this paradigm we suggest another conceptual picture, which is nearly as old as the "particle in a box" view of the electron (see Ref. 1), suggesting that the solvated electron is, in fact, an unusual kind of the solvent multimer anion in which the excess electron density occupies voids and cavities between the molecules in addition to frontier p-orbitals in the heteroatoms in the solvating groups. We argue that such a view does not contradict the experimental observations for the ground state electron and, in fact, accounts for several observations that have not been rationalized yet using the one-electron models, including the dynamic behavior of the excess electron. The emerging picture of the solvated electron is complementary to the familiar one-electron models, retaining and rationalizing the desirable features of the "particle in a box" paradigm and adding new features that are lacking in this class of models. In this respect, the multimer anion picture is different from more radical suggestions [61] that postulate a different *atomic* structure for the "solvated electron."

## 2. THE CAVITY ELECTRON

In the standard picture of electron solvation in polar liquids, the s-like excess electron occupies a (on average, nearly isotropic) solvent cavity that is stabilized through (i) Pauli exclusion of the solvent molecules (repulsive interaction) by the electron filling the cavity and (ii) point-dipole attractive interactions with the polar groups (such as HO groups in water) of 4-8 solvent molecules that collectively localize and trap the electron inside the cavity. Only the species in which some electron density is located inside this cavity (or the interstitial voids between the solvent molecules) can be rightfully called the "solvated electron." All such species exhibit a characteristic broad, asymmetric absorption band in the VIS (visible) or NIR (near infrared) most of which is from s-p excitation of the s-like ground state electron to three nodal p-like (bound) excited states (for electrons in deep traps in polar solvents) or free p-waves in the conduction band (CB), in less polar and nonpolar solvents. Since the cavity is slightly anisotropic, these p-like states are nondegenerate, and the VIS-NIR band is a superposition of three homogeneously broadened p-subbands. The more anisotropic is the cavity, the greater is the energy splitting between the centroids of these three subbands. To the blue of this composite s-p band, there is usually a Lorentzian "tail" extending towards the UV, due to the transitions from the ground state directly into the CB of the liquid. Some other excitations (s-d) might also contribute to this "tail" absorbance, according to the theory. [44-46,62] So characteristic is this "a bell and a tail" composite spectrum that most of the "solvated electrons" in liquids have been identified by this feature alone. Another distinguishing property of the solvated electron is pronounced temperature dependence of this spectrum, [13,14] with systematic red shift of the absorption peak and broadening of the absorption line with the increasing temperature. Solvated electron is, in fact, one of the best molecular "thermometers" in chemistry. These trends are commonly rationalized as an increase in the volume occupied by the electron as the cavity expands due to weakening of bonds between the solvent molecules, although there are experimental observations (such as the lack of spectral shift for $e^-_{hyd}$ in supercritical water as the density changes from 0.1 to 0.6 g/cm$^3$ [13] and much greater sensitivity of the absorption maximum to the changes of density that are induced by pressure than temperature decrease) that hint at the complex nature of the change observed that is not captured by the existing models of electron hydration. In particular, the recent suggestion that the energetics of solvation is solely a function of water density [12] does not appear to be supported experimentally. [13,14]

Negative EA is not a sufficient condition for the formation of the cavity electron: [10] the dimers and multimers of the solvent molecules should also have negative EA, lacking a way of accommodating the electron through the formation of bonded or stacked multimer anions. The depth of the potential well in which the electron resides broadly correlates with the solvent polarity. For such solvents as liquid water and ammonia, the trap is more than 1 eV below the CB of the liquid and thermal excitation of the electron into the CB is impossible. Solvated electrons in such liquids move adiabatically, following fast molecular motions in the liquid; in low-temperature solids, such electrons undergo trap-to-trap tunneling in competition with deepening of the traps due to the relaxation of the cavity, which may take as long as micro- and milliseconds at 20-100 K. In nonpolar liquids, some of which (e.g., alkanes) also yield solvated electrons, the traps are just 100-200 meV below the CB and thermal excitation of the electrons localized in such traps is sufficient to promote the electron back into the CB. Such electrons perpetually oscillate between the bottom of the CB and the solvent traps. Only recently has it been shown how fast are these electron equilibria in the alkanes: [2] the typical residence time of the electron in a trapped state is under 10 ps (at 300 K) and the typical trapping time of the quasifree CB electron is just 15-30 fs, which is shorter than the relaxation of its momentum. Note that the p-like electrons in polar solvents are also very close to the mobility edge of the liquid [41-43] and might be spontaneously emitted into the CB as a result of solvent fluctuations. [63] There is also a class of solvated/trapped electrons that coexist either in two distinct forms (such as electrons in low-temperature hexagonal ice [64,65] and cation-bound electrons in concentrated ionic solutions and glasses) [9,64,65] or in a dynamic equilibrium with a molecular anion (a monomer or a bound multimer), which occurs in some liquids (benzene, [66] acetonitrile [67]) or in dilute solutions of polar molecules in nonpolar liquids (e.g., for clustered hydroxylic molecules in dilute alkane solutions [5]). Many modes of electron localization are known and perhaps even more are still unknown.

For molecules lacking permanent polar groups (e.g., alkanes) or having the dipole oriented in such a way that the positively charged end of the dipole is looking outwards (e.g., ethers, amides, esters, and nitriles) the electron is solvated by the alkyl groups, and the trapping potential originates chiefly through the polarization of C-C and C-H bonds. [2] The exact origin of this potential is poorly understood; it appears that polarizability of these bonds, in the absence of percolation of the electron density onto the aliphatic chains, is insufficient to account for the energetics observed. The likely mechanism for electron trapping in such liquids, in the framework of multielectron approach, is examined in Refs. 2 and 4.

## 3. EXCITED STATES, PRECURSORS, DYNAMICS

### 3.1 "Hot" s-like and p-like states

For polar molecules with XH groups (X=N, O), the origin of the trapping potential is well understood: it is electrostatic interaction of the s-like electron residing inside the cavity and dangling (or "non-hydrogen-bonding") XH groups at the wall of the cavity; the wavefunction of the electron "washes" the protons in such groups and instantly responds to their rapid motion. The migration of the electron occurs either as the result of such molecular motions (in liquids) [68], tunneling to neighboring voids appearing as the result of solvent fluctuations, or repeated thermal emission to CB and relocalization. This relocalization can be induced directly by 1- [63] or 2-photon [31,32] excitation of the electron into the CB, or even by photoexcitation of the s-like electron into the p-like excited state, as the manifold of these p-states is close to the CB. [63] As for the existence of other than s- and p-like cavity states (such as 2s and d-states), while these states are periodically invoked in the literature (e.g., for vitreous ethers) there seem to be no recent corroborations of their existence, except for the electron bubbles in liquid $^4$He. Importantly, these are all virtual states: once the electron is excited into one of these states, many new states appear as the solvent accommodates to the excitation. These occupied states are classed into p- and s-state manifolds, although such a classification is somewhat misleading, as the wavefunction of the lowest ("s-like") state in an anisotropic, fluctuating cavity has substantial p- and d- characters. [44-46] The best understood of these excited states of the cavity electron are the so-called "hot" s-like states that relax adiabatically to the fully thermalized, fully solvated s-like state by damping their excess energy into the solvent. These are, basically, s-like states that are structurally and electronically very similar to the ground state of the electron but reside in a slightly modified cavity. Such states are produced in all situations when the electron is excited or ejected into the liquid preceding the formation of a fully thermalized s-like electron on the picosecond time scale. The spectral manifestation of the relaxation for these "hot" states is the so-called "continuous blue shift" of the s-p band that occurs on the time scale of 300-1000 fs in water [24-29,69,70] and even slower in alcohols and diols (a few picoseconds [25,71,72] or even a few tens of picoseconds, [29] in cold liquids). Only this type of dynamics was observed in photoionization of liquid water and electron photodetachment in the course of charge transfer to solvent (CTTS). [30] We have recently shown [28] (Fig. 1) that this process is bimodal: there is a rapid blue shift of the entire spectrum on the sub-picosecond time scale (that is not conserving the shape of the spectrum, contrary to the frequently made assumption [24-27,69-72]; Fig. 1a) and a slower narrowing of the spectral envelope to the red of the maximum that looks similar to vibrational relaxation in photoexcited molecules. During this delayed narrowing, the position of the absorption maximum is "locked" within 20 meV from the equilibrium position (Fig 1b). The time constants for this narrowing are 560 fs for $H_2O$ and 640 fs for $D_2O$, whereas the time constant for the initial blue shift is < 300 fs. Inability to distinguish between these two regimes, due to sparse spectral sampling and reliance on prescribed spectral evolution, might explain considerable scatter of time constants for "continuous blue shift" in the literature. Typically, such a shift (with the conserved envelope of the spectrum plotted as a function of excitation energy) is postulated rather than observed directly, as there are interfering absorbances and side processes occurring on the same time scale. Interestingly, the "hot" s-like electrons generated by 2 x 6.2 eV photoionization exhibit a short lived (< 100 fs) absorbance in the region where $H_2O$ has the third overtone of the O-H stretch mode (the 1.2 mm band in Fig. 1a that is not seen for heavy water) indicative of a strong vibronic coupling between the short-lived, energetic "hot" s-like state and the solvent. [28]

Place Figure 1 here

Apart from these "hot" s-like states, other light-absorbing states were identified using decomposition analysis of transient absorption spectra. There are as many such spectral/kinetic decompositions as there are authors, and relatively few common, agreed upon features have emerged from such analyses. The most likely culprit is the core assumption made in these decomposition analyses that a small number of states with well-defined, time independent spectra (or a species with prescribed "continuous blue shift" spectral evolution) suffices to account for the observed dynamics. The validity of such an assumption is unobvious for a species like solvated electron that is a statistical average over many solvent configurations. Only few general remarks are thus possible. There appears to be no evidence that p-like states are generated as detectable intermediates in the course of ionization or electron photodetachment, though such states may be generated by s-p and s-CB excitation of solvated electrons and large water anion clusters. The most likely reason for that is the extremely short lifetime of these p-like states (see below). In bulk liquids, these p-like states are predicted to exhibit diffuse p-CB absorption bands centered at < 1 eV. [73] The initial relaxation of these p-like states in bulk water is expected to occur very rapidly (10-30 fs); the inertial component of this relaxation is very pronounced and it is expected to exhibit a large isotope effect. For medium size water anion clusters in the gas phase, such p-like states are readily identified using angle-resolved photoelectron spectroscopy, as the photoelectrons generated from these states carry orbital momentum. [16,17] Two general schemes were put forward for the subsequent dynamics of the "relaxed" p-like states: (i) relatively slow adiabatic internal conversion (IC) and (ii) very fast nonadiabatic IC. In both of these scenarios the p-like states convert to a "hot" s-like state that subsequently undergoes adiabatic relaxation. In the adiabatic IC scenario, the lifetime of the relaxed p-like states is 100-300 fs [31-34]; this time increases to ca. 2 ps for methanol. [36] In the rapid, nonadiabatic IC scenario, this lifetime is on the order of 50 fs, and the 300-400 fs component is interpreted as the initial stage in the thermalization of the "hot" s-like state. [11] For $n$=25-50 water anion clusters, $\left(H_2O\right)_n^-$, the time constant for IC scales as $n^{-1}$ decreasing with the increased cluster size $n$ from 180 to 130 fs for $H_2O$ and 400 to 225 fs for $D_2O$. [16] Extrapolating these estimates to water bulk ($n \to \infty$) suggests that the time constant for IC is < 50 fs. While the validity of such extrapolation may be questioned [16,17,74], as it is not even fully established that such clusters trap the electrons internally, the recent measurements of emission lifetime of the p-like states in liquid water also give an estimate of 30 fs. [18] Photon echo and resonant transient grating measurements of Wiersma and co-workers [11] using 5 fs pulses suggested 50-70 fs time scale for the IC. These ultrafast measurements indicate the involvement of 850 $cm^{-1}$ libration mode in the solvent dynamics of photogenerated p-like states; the same modes show the largest resonance Raman enhancements. A large isotope effect of 1.4 on the lifetime of the fast (35 fs) component in the kinetics observed after s-p excitation by Barbara *et al.* [31,32] also implicates the involvement of these libration modes in the relaxation or IC of the p-like electron. The short lifetime of these p-like states readily explains the absence of the expected p-CB absorbances in various photoionization and photodetachment experiments and the observation of Assel *et al.* [34] that the same photoinduced absorbances are generated in the s-p and the s-CB photoexcitations: the transient absorption spectra are always dominated by the "hot" s-like states.

Nonadiabatic MQC MD calculations of Schwartz and Rossky [73] for the $e^-_{hyd}$ are consistent only with the slow adiabatic IC scenario, and only approximately, predicting much longer adiabatic relaxation of the p-like states (300 fs) and IC of relaxed p-like states (1 ps) and yielding unrealistically rapid relaxation of the "hot" s-states (<< 100 fs). There is still no satisfactory formulation of a dynamic MQC theory rationalizing rapid, nonadiabatic IC suggested by the experimental observations. It is likely that such a theory would soon be suggested: there is much activity in developing the next generation of MQC MD models (e.g., Ref. 52) that would be capable of addressing this issue. Very recently, Borgis, Rossky, and Turi, [53] re-estimated nonadiabatic IC rates using a kinetic model based on modified Fermi golden rule with either classical or standard quantized version of the correlation functions and obtained IC lifetimes of 30-60 fs for water and 80-160 fs for methanol. For harmonic quantization, even shorter p-like state lifetimes (a few fs) were computed. The authors caution that their results are very sensitive to the choice of model potentials; nevertheless, it appears that the time constant for IC would be short in models with more realistic treatment of electron-solvent interaction and quantized vibrations. The semicontinuum solvation model of Fischer and Zharikov [75] also favors short IC times (ca. 130 fs for water). For internally trapping an "octahedral" $S_6$ (n=6) water anion cluster, Scherer *et al.* [62] estimated an IC time of ca. 100 fs. It seems that the case for rapid, nonadiabatic IC for $e^-_{hyd}$ is growing stronger, whereas no new models or experiments favoring the slow, adiabatic IC scenario are appearing. If the former scenario is correct, transient absorbances that were attributed to p-CB bands (with the life times of a few 100s of ps) in Refs. 31-34 should be reinterpreted as those originating from the s-p bands of "hot" s-like states undergoing the first stage of their bimodal relaxation dynamics.

### 3.2 "Weakly bound" and "dry" electrons, relocalization, attachment.

Apart from these s- and p-like states, other short-lived, excited states were postulated by various authors. Such states go under the vague name of "weakly bound" (wb) electrons meaning a state that is localized yet not completely solvated; [24,25,72] it is assumed that such wb electrons in some way are structurally different from the strongly bound (sb) electrons observed at a later stage of the solvation process (the "hot" and the relaxed s-like states). In laser photoionization of water, some of these intermediate wb states (with lifetime of ca. 110 fs) were claimed to be the electrons coupled to the OH radical (yielding large transient absorbances near the O-H stretch of the water), [27] though no such states have been observed in electron photodetachment from hydroxide anion. [76] For water, there is no evidence that these wb electrons are more than artifacts of multiexponential kinetic analyses (see discussion in Ref. 28). [1] What is certain, however, is that the localization of the electron is preceded by the formation of short-lived *delocalized* states capable of extremely fast scavenging reactions with certain solutes (such as $Cd^{2+}$ and selenate and nitrate anions) [79-81] that are also known as scavengers of "dry" electrons generated in radiolysis of water. The evidence for such states is indirect (there is no spectroscopic signature); nevertheless, the existence of these states can be deduced from the occurrence of long-range scavenging (electron attachment) that occurs on the time scale < 50 fs. The typical static scavenging radii for s-like and p-like electrons by such solutes are 0.5 and 0.8-1 nm, respectively; whereas the CB electrons generated via s-CB excitation have radii of 3-5 nm. [63,79-81] Beyond these estimates, little is known about the nature of "dry" or CB electrons in polar liquids. The recent ultrafast photoemission studies of amorphous ice on metals suggest extremely short lifetime (< 20-50 fs) and rapid scattering for these CB states. [82,83] By contrast, CB electrons in low temperature ice-Ih are readily observed using dc and GHz conductivity, over many nanoseconds. [84]

In liquid ethers, such as THF, Schwartz and co-workers [37,54,85] observed theoretically, using nonadiabatic MQC MD model, a relocalization of photoexcited s-like electrons that proceeds via the formation of disjoint and multicavity states. Relatively large cavities occur naturally in these poorly packed liquids as the result of solvent fluctuations. The interaction of the solvent molecules with the electron is so weak that these nascent cavities have comparable binding energies to the relaxed cavities that are already occupied by the excess electron. [54] The "tail" in the absorption spectrum originates from weak transitions from the ground state to such disjoint and multicavity states rather than s-CB transitions. The classification of such states into s- and p- is not particularly useful due to the great anisotropy of the solvation cavity and numerous crossings between the excited states. Even the excited states that remain localized in their parent cavity after their relaxation may fleetingly occupy the neighboring cavities. Experiments on relocalization of solvated electrons in THF [37] are consistent with the picture of population transfer into these disjoint cavities that occurs bypassing the CB of the liquid; both the experiment and the theory indicate that ca. 30% of photoexcited s-like electrons relocalizes into such states (in the experiment, this process competes with geminate recombination). [37,85] It is not known whether such a mechanism can operate in H-bonded, well-packed solvents such as liquid water and alcohols, but one cannot entirely exclude such a possibility, especially at high temperature.

[1] It should be stressed that multiphoton pump-probe studies are frequently carried out at high excitation density; this may result in a bulk thermal spike that considerably changes both the electron thermalization and geminate recombination dynamics. [77,78] It seems likely that irreproducible reports of unusual spectral features and exotic short-lived intermediates may be traced to the effects of such thermal spikes.

Packing of the solvent molecules in an organic liquid has another important effect on the dynamics of electron localization: there could be more than one type of electron-trapping cavity and the interconversion between such cavities could be relatively slow, especially in viscous liquids. In aliphatic alcohols, the relaxation of "hot" s-like states is much slower than in water (e.g., 12 ps for methanol [24,25]), and the wb electrons can be observed directly on the sub-ps time scale (recombining or converting to a "hot" s-like state on the picosecond time scale). Spectrally distinguishable IR-absorbing wb state is observed most clearly in ethylene glycol at 300 K; the spectral evolution is consistent with 2.5 ps decay of this wb state without relaxation to a "hot" s-like state. [72] Mostafavi and co-workers [72] suggested that wb electrons in the diols are trapped electron species that are partially solvated by methylene groups; the wb electrons are observed before the conformational dynamics allow the OH groups to arrange properly around the solvation cavity. Such a picture is also suggested by *ab initio* calculations for methanol clusters [86] indicating possible participation of methyl groups. This rationale implies that there should be no "wb electrons" in liquid water (indeed, no such species in liquid water is suggested by the existing theoretical models or reliably observed experimentally). The situation is different in low-temperature hexagonal ice where rotations of water molecules are hindered; the wb electrons with life time of several milliseconds are readily observed in the $D_2O$ ice below 20 K [64,65]. The most likely trap site for these wb electrons is the so-called positively vested water vacancy (with three dangling OH groups interacting with the electron). [84]

In polar solids, the existence of wb electrons is beyond doubt, being richly documented in pulse radiolysis studies of vitreous alcohols, water-alcohol glasses, and salt glasses at low temperature. [64,65] Arrested molecular motions in such solids, long-range tunneling, and trap-to-trap downhill hopping of the electron readily explain the dynamics of wb electrons observed in these low-temperature glasses. [65] In fact, the hypothesis that the IR-absorbing wb electrons are partially or entirely trapped by the alkyl groups have been suggested by Shida *et al.* 35 years ago, [87] and this hypothesis is supported by the observed correlation of the position of the IR absorption peak with the length of the aliphatic chains and the similarity of this band to the absorption band of trapped electrons in vitreous alkanes [65,66] and small clusters of alcohol molecules in alkane liquids. [5,6] Ultrafast laser studies of liquid alcohols and diols thus recreate the familiar features of electron dynamics in low-temperature glasses, albeit on a much shorter time scale. It seems, therefore, that obtaining detailed structural data on these long-lived (> 1 ms in n-propanol at 77 K [65]) wb states in such solids would be preferable to the more involved studies of essentially the same wb states on the picosecond time scale, in room temperature liquids.

The relatively unaddressed issue is the dissociative electron attachment (DEA) involving photoexcited solvated electrons. DEA (presumably, involving protonated phosphate groups in the sugars) involving low-energy electrons (a few eV) has been implemented in inducing irreversible DNA damage; [88] DEA in water (involving a short-lived electron precursor, such as subexcitation electron) has been suggested as the main source of prompt $H_2$ in radiolysis of aqueous solutions. [89] DEA involving photoexcited solvated electron was observed in solid and liquid alcohols [36] and reverse micelles [87] but not in neat water, where the $H^- + HO^\bullet$ resonances are relatively high in energy. Given that DEA involving "precursor" states has been postulated for many radiolytic systems in order to account for the prompt bond breaking observed in such systems, more studies of the DEA involving energetic electron states are merited, as presently there is no other way of accessing such states in bulk solvents.

Concluding this section, we note that though the main intermediates of electron solvation have been identified, many controversies remain. Phenomenological approaches in interpreting these dynamics resulted in proliferation of mutually exclusive kinetic schemes providing limited insight into the physics and structural aspects of the electron dynamics. In our opinion, subsequent advance in understanding these dynamics can only be made by direct comparison of the experimental dynamics with theoretical models. Unfortunately, the current state of these models does not allow such a comparison. In the following, we will focus on the ground state of the solvated electron, as better understanding of this ground state is the likely key to developing theoretical models that can make this direct comparison possible.

## 4. THE CAVITY ELECTRON REVISITED

Studies of electron solvation are popular with chemical physicists largely due to the perceived simplicity of the problem. The latter notion rests upon the mental picture of the solvated electron as a single quantum mechanical particle confined in a classical potential well: "a particle in a box." This picture was first suggested by Ogg in 1946 and subsequently elaborated by Cohen, Rice, Platzmann, Jortner, Castner, and many others. First such models were static, but in the mid-1980s it became possible to treat the (classical) dynamics of the solvent molecules explicitly using computer models and MQC MD and path integral approaches flourished. The current state-of-the-art dynamic models are the descendants of these one-electron models. Despite their great sophistication, such models still rest on the initial *ad hoc* assumption that the cavity electron and the valence electrons in the solvent molecules may be treated wholly separably. Yet this basic assumption is unobvious, and as such, it has been the subject of much debate in the late 1960s and the early 1970s (that was eventually resolved in favor of the one-electron approximation). Indeed, there are multiple experimental observations that are not accounted for by these one-electron models.

For example, the one-electron models incorrectly predict (even at a qualitative level) the Knight shifts in $^1$H and $^{14}$N NMR spectra of ammoniated electron, $e_{am}^-$ [1], and solvated electrons in amines (Sec. 4.1). The same problem arises in the explanation of magnetic (hyperfine) parameters obtained from $^2$H ESEEM spectra of trapped (hydrated) electrons in low-temperature alkaline ices. [23] The recent resonance Raman spectra of $e_{hyd}^-$ [18-20] also appear to be incompatible with the one-electron models, as all vibrational bands (including the HOH bend) undergo substantial downshift that indicates weakening of the bonds (Sec. 4.2). Surprisingly, *the experimental methods that provide the most direct insight in the structure of the solvation cavity appear to be the least compatible with the one-electron models.* The latter models, however, do capture the essential physics of electron solvation, given their historical success in explaining the absorption properties, the energetics, the dynamics, and the spectral evolution of the electrons. These two lines of reasoning suggest that the one-electron models adequately describe the electron wavefunction inside the cavity ("the particle") but err in their description of the electron wavefunction extending beyond the cavity ("the box"). Both the hyperfine constants (that is, the spin density in the solvent molecules) and the vibrational frequencies of the solvent molecules are the properties of "the box." The "dissenting" experimental results, therefore, indicate that electron solvation significantly modifies the properties of these solvent molecules, and this salient feature is not included in the current MQC models.

### 4.1 Magnetic resonance

The way in which this solvent modification occurs is suggested by the pattern of hyperfine constants for $e^-_{am}$ (which is one of the few solvated electron species sufficiently stable to obtain its NMR spectrum). The Knight shift $K_X$ of NMR lines is due to the contact Fermi (isotropic) hyperfine interaction of the excess electron with the magnetic nuclei (*X*) in the solvent molecules; it is the measure of spin density $|\phi_s(0)|^2_X$ in the *s*-type atomic orbitals centered on a given nucleus *X*: $K_X \propto \Sigma_X |\phi_s(0)|^2_X$. This shift can be converted into the sum $\Sigma_X a$ of isotropic hyperfine coupling constants (hfcc's) for all nuclei of type *X*. In ammonia, this calculation gives +110 G for $^{14}$N nuclei and -5.7 G for $^{1}$H nuclei. Given that the atomic hfcc for the electron in the N 2s orbital is +550 G, ca. 20% of the total spin density of the excess electron is transferred into these N 2s orbitals. [1] Even more is expected to be transferred to N 2p orbitals, and this accounts for the negative sign of proton coupling constants: the spin density in the hybridized H 1s orbital is negative due to the spin bond polarization involving the filled N 2p orbital; this inversion is typical of p-radicals. The negative sign of the isotropic hfcc for protons was demonstrated by dynamic nuclear polarization experiments and then confirmed by direct NMR measurements. Note that only positively valued constants are obtained in the one-electron models for the s-like state, as such models do not include spin bond polarization effects.

A similar situation exists for "hydrated" electron trapped in low temperature alkaline ice. It was $^2$H ESEEM studies of this species that prompted Kevan [90] to suggest the well-known octahedral model of electron solvation (the so-called "Kevan's model") in which the electron is stabilized through dipolar interactions with six HO groups pointing to a common center. Following the original interpretation of these ESEEM spectra by Kevan *et al.* [90] (which gave positively valued isotropic hfcc's for the protons), it was subsequently demonstrated that isotropic hfcc's of the protons at the cavity wall (roughly 0.2 nm from the center) are, in fact, negative: $a \approx$ -0.92 G. [91] These negative hfcc's hint at nonzero spin density in the O 2p orbitals of water molecules. Tight-binding *ab initio* models of small anion water and ammonia clusters suggested the same. [92] However, given the many approximations made in these models, these dissenting results were not given due consideration at a time when the conceptual picture of the solvated electron was still evolving. The consensual picture that emerged in the mid-1970s was that the solvated electron is indeed a "particle in a box," to a very good approximation. The competing view that the solvated electron is a multimer anion was suggested for $e_{am}^-$ by Symons, [93] who estimated that the spin density is divided between 6 ammonia molecules in the first solvation shell (with hfcc of 12 G) and 12 molecules in the second solvation shell (ca. 3 G). A more detailed accounts of magnetic resonance studies of ammoniated and hydrated electron given in Refs. 1 and 23, respectively. In retrospect, the main reasons for rejection of the anion picture of electron solvation (that has been around since 1953 [94]) was the insistence of the proponents that *all* of the excess electron density resides on the solvent molecules, their inability to explain the observed energetics and absorption properties of the solvated electron, and their denial of cavity formation.

The latter was especially damaging, as there is abundant evidence that such cavity electrons do exist in molecular fluids. [95] Furthermore, the existence of the cavity logically follows from the occurrence of charge sharing between several solvent molecules: although only a small fraction of the negative charge resides on each solvent molecule forming the cavity, Coulomb repulsion between these partially charged molecules assists in opening of the cavity. The latter can be formed through this mechanism even when the occupancy of the cavity is relatively small (as is the case in the alkanes [4]). The conflict with these incontrovertible experimental observations is resolved by making the assumption that only a fraction of the total negative charge is localized on the solvent molecules; the rest is localized inside the solvation cavity. This view, first suggested by Symons [93] and Kevan, [96] partakes of the best features of the cavity and the multimer anion models.

## 4.2 Vibrational spectroscopy

Magnetic resonance is not the only piece of evidence indicative of the excess electron density in the solvent molecules. The hydrated electron exhibits a second absorption band at 190 nm [97] that originates through the perturbation of O 2p orbitals in the solvating water molecules; obviously, such a feature cannot be treated using the standard one-electron models. Further evidence is suggested by the recent resonance Raman (RR) observations. [18-22] The vibrational peaks of $e^-_{hyd}$ (which demonstrate resonant enhancements over $10^5$) all exhibit significant downshifts relative to these Raman peaks in neat water. In RR spectroscopy, only those vibrational modes that are significantly displaced upon electronic excitation show resonance enhancement; thus, this spectroscopy provides a probe of the water molecules in the immediate vicinity of $e^-_{hyd}$. The RR peak position for the $e^-_{hyd}$ in $H_2O$ (vs. those for bulk water), in $cm^{-1}$, are: librations at 410 (vs. 425-450), 530 (vs. 530-590), 698 (vs. 715-766); the H-O-H bend at 1610 (vs. 1640); and the H-O stretches at 3100 (vs. 3420). [18,19] Thus, the downshift of the bend mode, which exhibits a narrow, symmetric line, is ca. 30 $cm^{-1}$, and the downshift of the stretch mode is 200-300 $cm^{-1}$. Similar RR downshifts were observed for solvated electrons in alcohols, with the downshift of the O-H stretching mode increasing with the solvent polarity. [21] For methanol, the O-H torsion peak is downshifted by 180 $cm^{-1}$ and the O-H stretch is downshifted by 340 $cm^{-1}$, which is greater than the downshift of the O-H stretch in liquid water. This large downshift is readily explained by the reduction in the number of solvation OH groups (from 4 to 6) in methanol, as suggested by ESEEM spectroscopy (see Ref. 23 for a review) which results in greater penetration of the electron density into the O 2p orbitals (see below). Normal mode analysis of enhancement factors for RR peaks indicates that all enhanced modes are predominantly O-H in character; the C-H stretching bands that have no OH character are not observed and the largest enhancements are for libration modes. So far, all RR observations are consistent with the notion that the electron in water and alcohols is solvated by dangling OH groups, and the electron-solvent coupling is mediated primarily by these OH groups; there is no evidence that sb electrons are solvated by CH groups of the alcohols. A more speculative idea suggested by Stuart *et al.* is that the alignment of these OH groups changes from pointing straight towards the center of mass (X) of the electron to dipole coupling to the alcohol molecule, as the carbon number increases from 1 to 4 (with the X-O-H angle increasing by $30^o$, respectively. [21]

These downshifts are such a general feature of the RR spectra that there must be a common mechanism for these downshifts in all solvents. Observe that the disruption of the H-bond structure cannot account for these RR results: weakening of the H-bonds results in O-H band upshifts, and the H-O-H bending mode in water does not change even when this liquid is heated to 600 K or saturated with salts (in fact, halide anion solvation in alcohols upshifts the O-H stretch by 50 $cm^{-1}$). [21] It has been concluded [18,19] that the only way of explaining these substantial downshifts is by assuming partial occupancy of solvent antibonding orbitals.

Most recently, Mizuno *et al.* presented a femtosecond version (250 fs time resolution, 160 $cm^{-1}$ spectral resolution) of the RR experiment to probe the O-H band of the electron as it hydrates following 2 x 4.66 eV photon excitation. Mizuno *et al.* [22] conclude that the precursor of the hydrated electron that undergoes "continuous blue shift" on the time scale of 1-2 ps also yields a downshifted O-H stretch signal whose resonance enhancement follows the efficiency of Raman excitation as the absorption spectrum of the s-like state shifts to the blue (thus indirectly confirming its identity as a "hot" s-like state). The comparison of anti-Stokes and Stokes Raman intensities indicates that the local temperature rise is < 100 K at 250 fs. This estimate agrees with the estimates based on the evolution of the spectral envelope during the thermalization, using the dependence of the absorption maximum of thermalized electron on the bath temperature. [26,69]

### 4.3 Substructure of the s-p absorption band

The s-p absorption band of the solvated electron can be thought of as three overlapping, homogeneously broadened s-p transitions. Variation of the Raman depolarization ratio across the O-H stretch band and significant deviation of this ration from 1/3 for the O-H stretch and libration bands indicate that the p-like states are nondegenerate (which does not exclude considerable homogeneous broadening of the lines). [18] The analyses of the spectral envelopes using RR data also suggest that inhomogeneous broadening is much stronger than homogeneous broadening, both for the water and the alcohols. [18,21] These observations are inconsistent with the recent suggestions of Wiersma and co-workers [11] that the electron spectrum is a single inhomogeneously broadened line (which is also in striking disagreement with all existing dynamic models of electron solvation). Still, at the present there is no further experimental evidence other than the RR results that distinct p-subbands do exist. Ultrafast laser experiments [34,35,98] that were specifically designed to demonstrate this subband structure using polarized transient hole burning (PTHB) yielded no conclusive evidence for this structure. PTHB spectroscopy is a form of pump-probe spectroscopy that examines the ground-state dynamics of a system by first exciting a subset of members of an ensemble with polarized light and then probing at a later time the dynamics of the remaining, unexcited members with light polarized parallel or perpendicular to the original excitation polarization. [35,99] If the three p-like states interchange roles slowly (as the solvation cavity deforms in response to solvent motion), then PTHB should show different dynamics for these two probe polarizations. When the lowest-energy transition along the long axis of the cavity is excited with polarized light, until the cavity reorients, there would be less absorption by the remaining electrons when probing with light of the same polarization at the excitation energy but the remaining electrons would continue to absorb at higher energies. The parallel and perpendicularly-polarized THB signals should become identical once solvent motions have scrambled the three p-like states and memory of which transition dipole moment pointed which direction in space is lost. MQC MD simulations of Schwartz and Rossky [99] predicted that pumping the lowest-lying transition and probing either the same or the higher-lying transitions should give an anisotropy that persists for 1 ps, as it takes this long for the water molecules to rearrange enough that the cavity changes shape and thus scrambles memory of the transition dipole directions. Reid *et al.* [98] reported this persistent anisotropy, but subsequent studies [34,35] did not confirm the presence of a long-lived anisotropy. Explaining this striking inability of the PTHB experiment to demonstrate the predicted effect remains an open problem, and solving this problem requires reexamination of the origin of the absorption spectrum and the dynamics of the $e^-_{hyd}$. The discrepancy may be due to strong homogeneous broadening of the absorption spectrum of the $e^-_{hyd}$ or extremely fast interchange of the p-like orbitals. [35] We suggest, [100] however, that the failure might be with the very concept of the excess electron as "a particle in a box."

**5. THE HETERODOXY: MULTIMER ANION PICTURE OF ELECTRON SOLVATION**

In a series of recent publications on the excess electrons in ammonia [1] and water, [23,51] we addressed these issues by developing multi-electron models of electron solvation and demonstrating how such models account for the known properties of the solvated electrons. Our models are inspired by recent *ab initio* and DFT studies of large, internally trapping water anion clusters. [47-49] What follows from these latter studies is that the electron is localized by several dangling OH groups; in the medium size water clusters ($n$=17-24), the cavity is typically tetrahedral. Examination of the highest occupied molecular orbitals (HOMO's) of these clusters suggests that part of the electron density is contained in the frontal orbitals of the dangling OH groups. Computations of Kim *et al.*[47] and Domcke *et al.*[61] suggest that small internally trapping clusters ($n$<10) have some of their vibrational bands downshifted with respect to neutral water clusters. Recent calculations of vibrational properties of water anion clusters that trap the electron externally, by dipole binding to the so-called AA (double acceptor) water molecule at the surface of the cluster, [101,102] suggest that the main cause for these red shifts is donor-acceptor stabilization between the unpaired electron and O-H $\sigma^*$ orbitals. [101] This is, basically, another way to describe the mechanism suggested by Tauber and Mathies for $e^-_{hyd}$ [18] and Symons for $e^-_{am}$. [93] It thus appears that most of the physics which is necessary to address the problems discussed in Sec. 4 is already contained in such *ab initio* and DFT models. The problem with this inference is that the small and the medium size water anion clusters in the gas-phase have quite different structure from the $e^-_{hyd}$ in the bulk water, and this makes direct comparison impossible. Furthermore, it is well understood that $e^-_{hyd}$ is a dynamic entity, a statistical average over many solvent configurations that cannot be adequately represented by any given structure; a quantitative description of the $e^-_{hyd}$ within the multielectron approach has to address this inherent variability. One path to this goal is by using Car-Parrinello molecular dynamics (CPMD), and such a calculation for the $e^-_{hyd}$ in the room temperature and supercritical water has been implemented by Boero *et al.* [50] Unfortunately, CPMD is a computationally demanding approach, and this requires the use of small solvent cells of a few tens of water molecules; the solvated electron fills the substantial part of these cells. Since the solvent cell also has net negative charge, diffuse positive charge has to be spread on the neighboring cells, further reducing the fidelity of the model. To speed up the computation, CPMD calculations involve pseudopotentials; the use of such pseudopotentials has to be justified. For these or other reasons, the results of the CPMD calculation [50] look quite different from both MQC MD calculations for electron in liquid water and *ab initio* and DFT calculations for gas-phase water anion clusters. To further complicate the assessment of the CPMD results, the computation of magnetic resonance parameters, distribution of charge, absorption spectra, and PTHB dynamics requires high-quality local expansion of the wavefunction which is difficult to achieve using plane wave sets that are used for the CPMD calculations.

Hence we suggested a different approach that is less computationally demanding but appears to successfully capture the essential physics of the problem. [51] Our approach capitalizes on the remarkable success of MQC models to account for the electron properties. [38-46] MQC MD was used to generate a dynamical trajectory of the s-like $e^-_{hyd}$, and then temporally well-separated snapshots from this trajectory (100 fs per frame) were extracted and became the input for DFT and single-excitation configuration interaction (CIS) calculations. In these calculations, only one or two complete solvation shells for the excess electron were considered explicitly; the remaining atoms in the simulated solvent were replaced by point charges, a procedure that is referred to as matrix embedding (this approach has been used to study neutral water [103] and hydrated radicals [104]). Significant sharing of spin and charge of the excess electron by O 2p orbitals in the first-shell water molecules was observed (ca. 20%). This hybrid MQC MD:DFT(CIS) approach can account for (i) the energetics and the equilibrium optical spectrum of the $e^-_{hyd}$ in the visible and the UV; (ii) the EPR and ESEEM spectra and (iii) the vibrational (resonance Raman) spectrum of the $e^-_{hyd}$. Although the true multielectron picture of the $e^-_{hyd}$ is complex, on average, the radial density of the HOMO and the three lowest unoccupied molecular orbitals (LUMO) resemble the s-like and p-like orbitals predicted by the one-electron models. For some observables (e.g., the optical spectrum), the fine details of this orbital structure do not matter. For other observables (e.g., the spin density maps provided by EPR and ESEEM spectroscopies and the resonance Raman spectrum), this level of approximation is inadequate.

Place Figure 2 here.

The typical HOMO of the hydrated electron is shown in Fig. 2 (on the left) next to the average radial wavefunction (on the right). While there is considerable density in the O 2p orbitals (with the negative frontal lobes accounting for 12% of the total density), the ensemble average wavefunction is hydrogenic. The most probable position of the electron is at 0.175 nm, which is within the cavity radius of ca. 0.226 nm (the mean distance between the center of mass

X of the electron and the nearest protons in dangling OH groups). The mean X-O-H angle is close to 16% so the OH groups are oriented towards the cavity center. Ca. 50-60% of the density is contained within the cavity, with only 5% contained beyond the first solvation shell. Mulliken population analysis indicates that the excess charge and spin densities are localized mainly on the H and O atoms in the dangling OH groups. The radius of gyration of the electron is estimated as 0.275 nm (vs. 0.204 nm in the MQC MD model and experimental estimate of 0.25-0.26 nm) and the semiaxes of the gyration ellipsoid (the measure of cavity anisotropy) are 0.15 nm x 0.16 nm x 0.17 nm. This anisotropy splits the energies of the lowest unoccupied orbitals. The computed DOS function (Fig. 3a) exhibits two features near the bottom of the CB. Feature (i) results from the HOMO (the s-like orbital) that is located ca. -1.69 eV below the vacuum energy (the DOS maximum is at -1.8 eV vs. -1.75 eV given by the CPMD calculation). [50] Feature (ii) derives from the three lowest unoccupied molecular orbitals, which have centroids at 0.42, 0.65, and 0.86 eV, respectively, that correspond to the three nondegenerate p-like states observed in one-electron models. The histograms of the corresponding transition energies show three distinctive p-subbands with centroids at 2.11, 2.34, and 2.55 eV (Fig. 3b). For comparison, path integral calculations [38-40] using the same pseudopotential as for our MQC MD calculations gave peak positions at 2.1, 2.5, and 2.9 eV. The absorption spectra calculated using the CIS method are very similar to those calculated using the MQC MD method. The three p-subbands correspond to the three lowest excited states that have nearly orthogonal transition dipole moments. Each one of these subbands carries an integral oscillator strength of ca. 0.3. In Fig. 4, isodensity contour plots of the Kohn-Sham LUMO is shown. The familiar dumbbell shape of the ‘p-like orbital’ is not readily recognizable, although the three lower unoccupied states do exhibit p-like polarization, each orthogonal to the others. Only a fraction of the total ‘p-like state’ density (ca. 20%) is contained inside the cavity and the p-character of these electronic states is achieved mainly through the polarization of the frontal O 2p orbitals in the OH groups forming the cavity: the phase of the electron in these orbitals on one side of the cavity assumes a positive sign, while the phase of the electron in the O 2p orbitals straight across the cavity in the direction of the transition dipole moment assumes a negative sign (Fig. 4). There is also both positive and negative excess electron density in the interstitial cavities between the water molecules of the first and the second solvation shells. The gyration ellipsoid for these p-like orbitals is 0.18 nm x 0.22 nm x 0.33 nm making them nearly twice the size of the gyration ellipsoid for the HOMO, so the 'p-like' states extend further out of the cavity than the ‘s-like’ ground state. This readily accounts for the experimental observation that s-p excitation of the electron causes relocalization, albeit with a low probability. [63] The complex orbital structure of the p-like orbitals may also be important for understanding the mechanism for rapid, nonadiabatic IC involving these orbitals (Sec. 3.1). Indeed, the orbital momentum carried by the polarized cavity should be rapidly lost via small-amplitude motions (librations?) of the water molecules.

Place Figures 3 and 4 here.

A novel feature that is not captured by one-electron models is a band of HOMO-1 orbitals that are composed of $1b_1$ orbitals ( O 2np orbitals) of the water molecules in the first solvation shell (Fig. 3a). Our calculations suggest that the onset of this band starts 5.75 eV below the vacuum level. The presence of this peak suggests that there should be an electronic transition from the occupied O 2p orbitals into the HOMO at ca. 210 nm. The experimentally observed UV band of the $e_{hyd}^-$ peaks at 190 nm with an onset around 220 nm. By placing the unit negative charge at the center of the cavity, one can demonstrate that this feature originates through a Stark shift of the eigenvalues towards the midgap, by ca. 1.1 eV.

Using DFT calculations, it is possible to calculate hyperfine constants and then simulate $^{1}H$ EPR and $^{2}H$ ESEEM spectra of the $e^-_{hyd}$. The correspondence between such simulated and experimental [91] spectra is very good, with all of the salient features discussed above reproduced. The residual discrepancy is for $^{17}O$ nuclei: the calculated second moment ($M_2$) of the EPR for the 37% oxygen-17 enriched sample studied by Schlick *et al.*, [105] is 2250 G$^2$ vs. the reported experimental estimate of 134 G$^2$. This is not a failure of the particular DFT model: all *ab initio* and DFT models of the $e^-_{hyd}$ give large estimates for isotropic hfcc's on oxygen atoms. The estimate of Schlick *et al.* [105] is compromised by their subsequent observation [106] of a strong spectral overlap between one of the resonance lines of the $^{17}O^-$ radical and the narrow EPR signal from the "electron," which had a peak-to-peak width ($\Delta B_{pp}$) of 18±1 G. In $^{16}O$ glasses, the two narrow EPR signals from $e^-_{hyd}$ and $^{16}O^-$ are spectrally well separated, but because the signals overlap in $^{17}O$ enriched samples, the EPR spectrum in such $^{17}O$ enriched samples is very complex. We used our calculated hfc tensors to simulate the EPR spectrum of an oxygen-17 enriched sample. The EPR line decomposed into two distinct spectral contributions, a narrow one with $\Delta B_{pp} \approx 23$ G and $M_2 \approx 135$ G$^2$ (in good agreement with the estimates of Schlick *et al.* [105]) and a very broad line with $\Delta B_{pp} \approx 89$ G and $M_2 \approx 1980$ G$^2$. For a sample with 37% $^{17}O$ enrichment, there is a ca. 10% probability that the first solvation shell would have no $^{17}O$ nuclei. The narrow line arises from such isotopic configurations, so that the electron is only weakly coupled to the $^{17}O$ nuclei in the second solvation shell. The isotope configurations that include at least one $^{17}O$ nucleus in the first solvation shell, on the other hand, are responsible for the broad line. Small-amplitude movements of water molecules in the frozen samples would cause efficient spin relaxation for this line. The narrow EPR line was recognized as a signal originating from $e^-_{hyd}$ from its long relaxation time, using microwave saturation of the resonance signals. Broad resonance lines were attributed to the $^{17}O^-$ radical; this criterion eliminates strongly coupled water anion configurations. We believe, therefore, that the EPR results for oxygen-17 enriched samples do not contradict the MQC MD:DFT model.

To examine the vibrational spectra of the $e^-_{hyd}$, we calculated IR and Raman spectra of the embedded clusters retaining only the first solvation shell, as such calculations do not include resonant enhancement. Although the absolute positions of the vibrational features calculated using the embedded neutral water and water anion clusters do not match experiment, the downshifts of these bands in the presence of the excess electron are well described by the hybrid calculation: there are downshifts of the librational, the H-O-H bending and the O-H stretch modes. The calculated downshift for the H-O-H bending mode is ca. 50-60 $cm^{-1}$ (as compared to the experimental estimate of 30 $cm^{-1}$ [18]) and the calculated downshift for the O-H stretching modes is 80-180 $cm^{-1}$ (vs. 200-300 $cm^{-1}$ for the band center). [18] Electrostatic interactions alone cannot account for these downshifts, as placing the unit negative charge at the center of the cavity does not cause such large shifts.

Interestingly, the multielectron model also accounts for the "failure" of the PTHB experiment: since part of the transition dipole moment is carried out by O 2p orbitals in water molecules, their rapid reorientation quickly destroys the correlation. The slow reorientation of the cavity is still observed, but it is predicted to have little weight in the corresponding correlation functions. Direct calculation of PTHB dynamics using the method used by Schwartz and Rossky [99] and the transition dipole moments calculated using the CIS model indicates that (i) the anisotropic PTHB signal is very small (less than 2-5% of the isotropic contribution) and (ii) this signal fully decays in 250 fs. As the earliest delay time at which the PTHB signal is observed is ca. 200 fs, [35] the lack of the signal is readily rationalized.

For ammonia, [1] the DFT calculations suggest that small clusters that exhibit internal trapping of the electron automatically yield large positive Knight shifts on $^{14}N$ nuclei and small negative Knight shifts on $^{1}H$ nuclei. In a typical $n$=18 cluster, the spin density was mainly contained between three dangling NH bonds, but the diffuse s-like orbital enveloped the entire cluster. The isotropic hfcc's for $^{14}N$ nuclei of the three nearest molecules are 16-20 G. The "second solvation shell" molecules have hfcc's ranging from +1.9 to +6.6 G, depending on the proximity to the central cavity. The sum totals of isotropic hfcc’s for $^{14}N$ and $^{1}H$ nuclei are $\Sigma_N a \approx +117$ G and $\Sigma_H a \approx -4.1$ G, respectively, in good agreement with the experiment. These calculations strengthen the case for extensive sharing of spin density by N 2p orbitals by ammonia molecules in the first and the second solvation shells.

Even more extensive delocalization might occur for solvated electrons in alkane liquids. In one-electron models, the degree of the delocalization is determined by the binding energy of the electron. In alkanes, where the binding energy is 50-200 meV, the s-like electron (even in the one-electron models) spreads well beyond the hard core radius of the cavity. [2,4] This is in contrast to the electron in water that is still largely confined inside the solvation cavity. Thus, the problem of adequate description of the interaction of the electron and the solvent molecules is even more important for the alkanes. In our recent study, we demonstrated that the most likely way in which the nitriles and the alkanes "solvate" the electron is through the formation of a cavity in which the electron is mainly contained in the C 2p orbitals of the methyl (methylene) groups forming the solvation cavity. [4] For alkanes, the spin density spreads along the aliphatic chain, with alternating occupancy of C 2p orbitals receding towards the ends of the chains removed from the cavity. Such a structure for the excess electron in the alkanes further erases the distinction between the solvent stabilized multimer radical anion and the cavity electron.

One can ask a question, does such a distinction exist at all in the solvents of low polarity? The same question can be brought to a focus by the following thought experiment ("electron encapsulation"). [3,4] Suppose that the entire first solvation shell of the solvated electron is replaced by a single supramolecular structure (the "cage") that has the internal cavity lined by polar groups. The cage is suspended in a liquid with low binding energy for the excess electron. Assuming that the cage traps the excess electron, what is the result of this capture? Should one regard the resulting species as a "solvated electron" or as a molecular anion? We have recently addressed this problem experimentally [4] and theoretically, [3,4] for hydrogenated calixarene and polynitrile rings, and the answer appears to be that no firm criteria exist for classification of such borderline species. Multimer solvent anions (electron residing on the molecules) and cavity electrons (electrons residing in the voids between the molecules) are two realizations of the same structural motif; the real "solvated/encapsulated" electron is always in between these two extreme cases.

**5. CONCLUDING REMARKS**

To conclude this review, despite rapid progress, many outstanding questions about the solvated electron remain unanswered. The structure and the behavior of these unusual species turned out to be much more complex than originally believed. Further advances will require greater focus on the quantum chemical character of the "solvated electron" explicitly treating the valence electrons in the solvent and more realistic dynamic models of the solvent degrees of freedom and electron-solvent interactions. Developing a many-electron, dynamic picture of the "solvated electron" presents formidable difficulty, yet this is a task that cannot be avoided, as the potential of one-electron models to address the problems is inherently limited. The simplicity of the solvated electron (that is its major attraction to chemical physicists) is imaginary; the solvated electron is an orbitally complex, nanoscale solvent multimer anion. The ideal object imagined by Ogg nearly 60 years ago, "the particle in a box," has distant relation to the species observed experimentally.

Below we provide a short list of the important problems concerning the solvated electron in polar media (author's choice, no particular order):

- Why is the lifetime of p-like states of hydrated electrons so short? What is the structure of these p-like states? Are there other cavity states in water? disjoint states? multicavity states? How to prove their (non)existence experimentally?

- Can the electron be "simplified" by restriction/removal of the solvent degrees of freedom? Can it be encapsulated?

- How does the nature of the solvated electron changes from one liquid to another? Does it become more of a multimer anion as the polarity decreases? How does one describe the dynamic behavior of such a multimer anion species?

- What is the nature of wb ("weakly bound") electrons observed on the short time scale? Does such a species exist in liquid water? Are the wb electrons in alcohols partially solvated by their alkyl groups? Are these the same species that are observed in low-temperature solids?

- How do photoinduced relocalization and photoejection of the electron occur? Is there indeed a "conduction band" in polar liquids? Does the light-induced relocalization of the electron involve this "conduction band?" What kind of species is the "dry" electron? Is there actually such a species? Can it be an excitonic state of the solvent?

- How does water vibrate around the hydrated electron? What is the effect of these vibrations on the absorption, electronic, and dynamic properties of the solvated electron? What is the mechanism for relaxation of "hot" s-like electron on the sub-picosecond and picosecond time scales?

- How do water anion clusters in the gas phase relate to the solvated electrons observed in the bulk? How does 2D electron localization in layers of polar molecules on metal and metal oxide surfaces [83] relate to 3D localization in the bulk?

- What is the structure and the dynamics of hydrated/solvated electron in hot/supercritical water? in dispersed clusters of polar liquids in nonpolar liquids? in microheterogeneous media (e.g. water clusters in zeolite cavities)? in mixed solvents? on surfaces?

- Can the solvated/trapped/encapsulated electron be used for molecular electronics and quantum computing? It is the tiniest capacitor known in chemistry and the electron degrees of freedom are largely decoupled from the nuclear ones. Can the solvated electron be the organic chemistry substitute for quantum dots?

We hope that this review will foster interest in these problems. This work was performed under the auspices of the Office of Basic Energy Sciences, Division of Chemical Science, US-DOE under contract number DE-AC-02-06CH11357. The author thanks M. C. Sauer, Jr. for technical assistance and R. Mathies and D. M. Bartels for useful discussions and communicating to the author prepublication versions of their recent papers.

**Figure captions.**

**Figure 1.**

Spectral evolution of the "hot" s-like state of hydrated electron generated in two 6.2 eV photon ionization of light water. The arrows indicate the trends observed in the direction of longer delay times of the probe pulses. Panel (a) demonstrates the evolution between 500 fs and 1.2 ps, showing considerable blue shift and fast decay of the IR features. Panel (b) shows the slow relaxation regime that is observed after 1.2 ps (note the logarithmic scale). In this regime, the band maximum is "locked" within 20 meV and the spectral evolution is due to relatively slow, isotope sensitive narrowing of the spectral envelope on the red side of the spectrum. This narrowing is likely to be caused by vibrational relaxation of the "hot" s-like state. See Ref. 28 for more detail.

**Figure 2.**

(On the left) Isodensity map for s-like HOMO of hydrated electron given by MQC MD - DFT calculation; a single snapshot is shown (only two solvation shells are shown, the embedding matrix of water molecules is removed for clarity). The central s-like orbital (grey) has the opposite sign to frontier O 2p orbitals in water molecules "solvating" the electron. Ca. 20% of the electron density is in these O 2p orbitals. Despite that, the ensemble average radial component of the HOMO (on the right, solid line) closely resembles hydrogenic wavefunction (broken line). On average, ca. 60% of the electron density is contained inside the cavity and 90-95% within the first solvation shell. See Ref. 51 for more detail.

**Figure 3.**

(a) Kohn-Sham density of states (DOS) function for "hydrated electron" (embedded water anion clusters). The three core orbitals of water are shifted by 1 eV towards the midgap as a result of Stark shift. Features (i) and (ii) originate from s-like HOMO and p-like LUMO(0,+1,+2) orbitals. (b) The histogram of energy gaps between the s- and p-like states.

**Figure 4.**

The anatomy of a 'p-like state'. Two isodensity contour maps (±0.01 and ±0.03 a.u.$^{-3}$) of the same LUMO orbital are shown side by side. Unlike the p-like orbitals in one-electron models, LUMO states in MQC MD - DFT and CIS models have the lobes pushed outwards between the first and the second solvation shells, with $< 20\%$ of the spin density residing inside the cavity. This results in considerable fragmentation of the diffuse part of the wavefunction. The O 2p orbitals are strongly polarized, with opposite signs of the orbitals attained by water molecules on the opposite sides of the cavity in the direction of transition dipole moment.

Figure 1. Shkrob

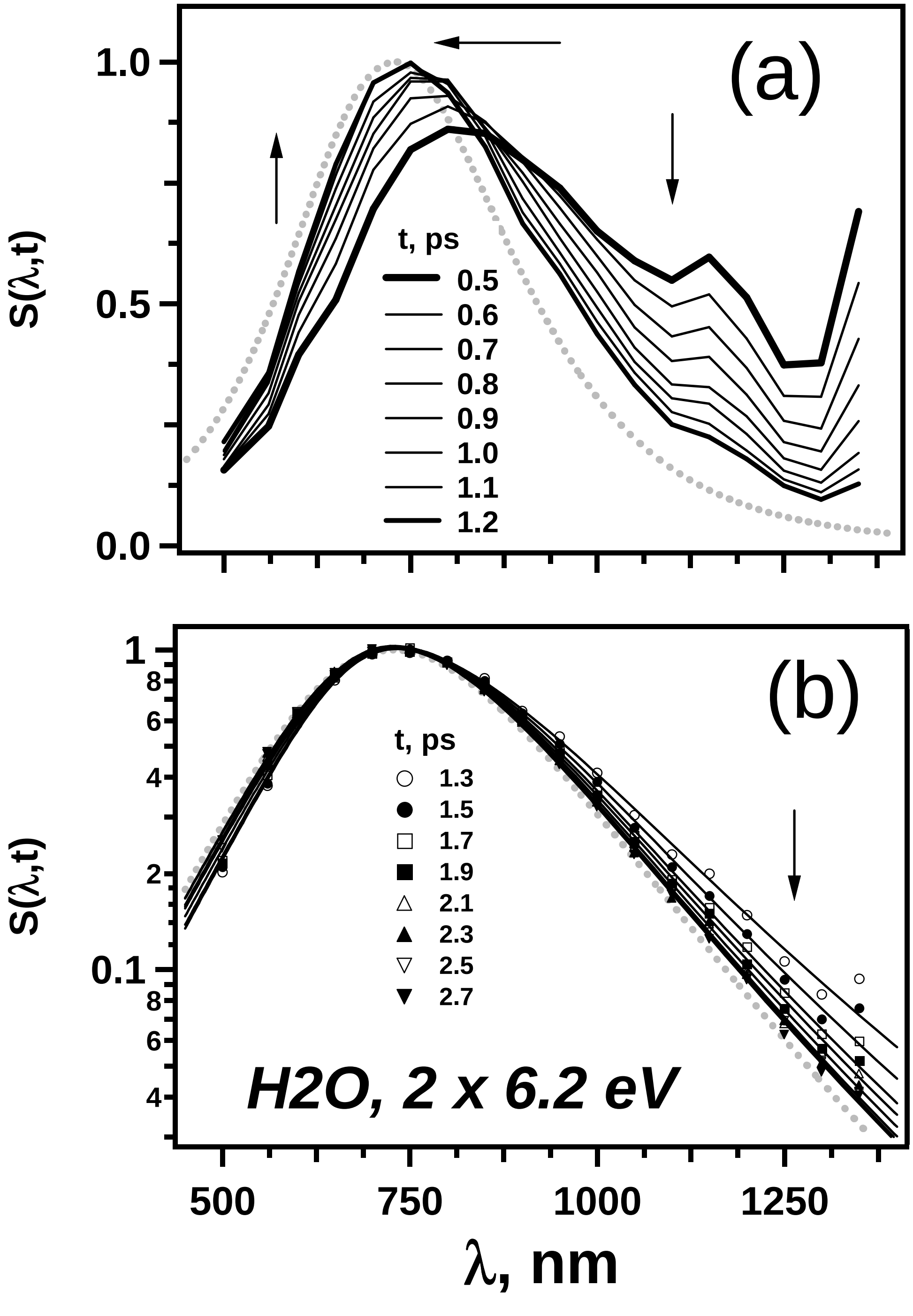

Figure 2. Shkrob

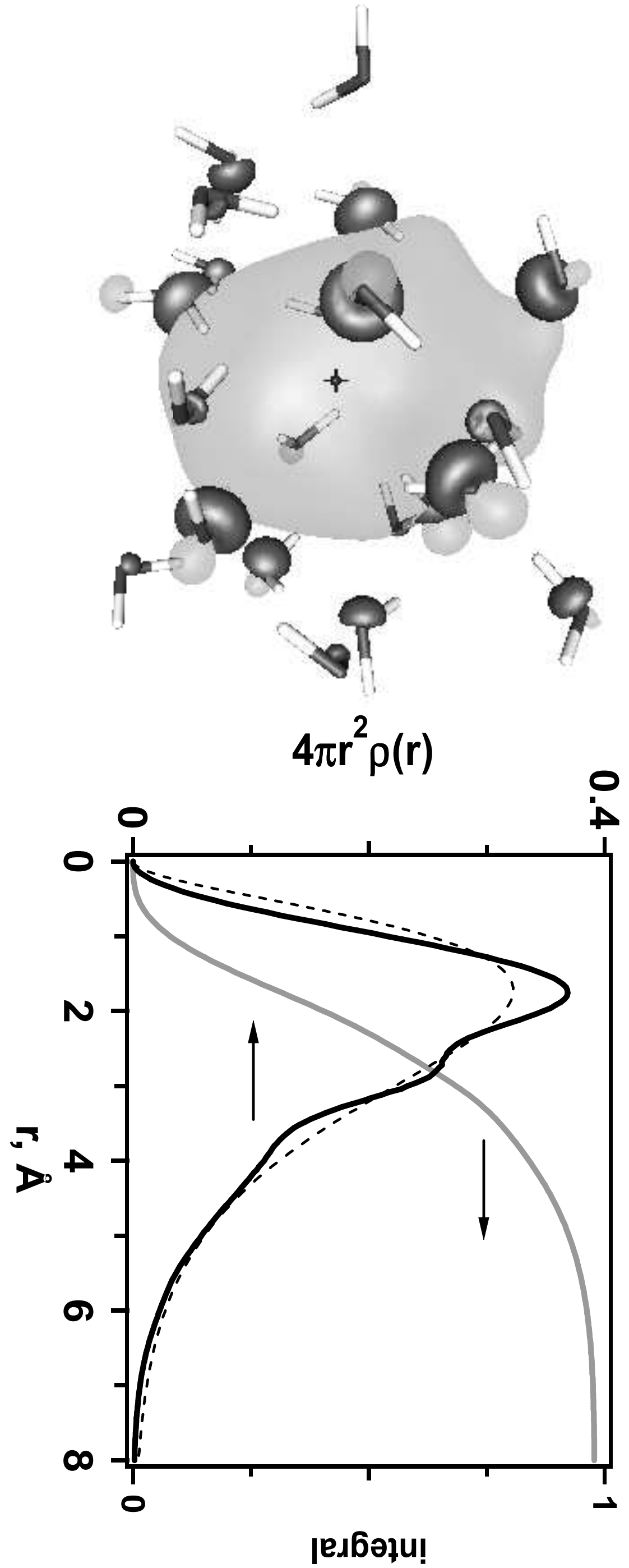

Figure 3. Shkrob

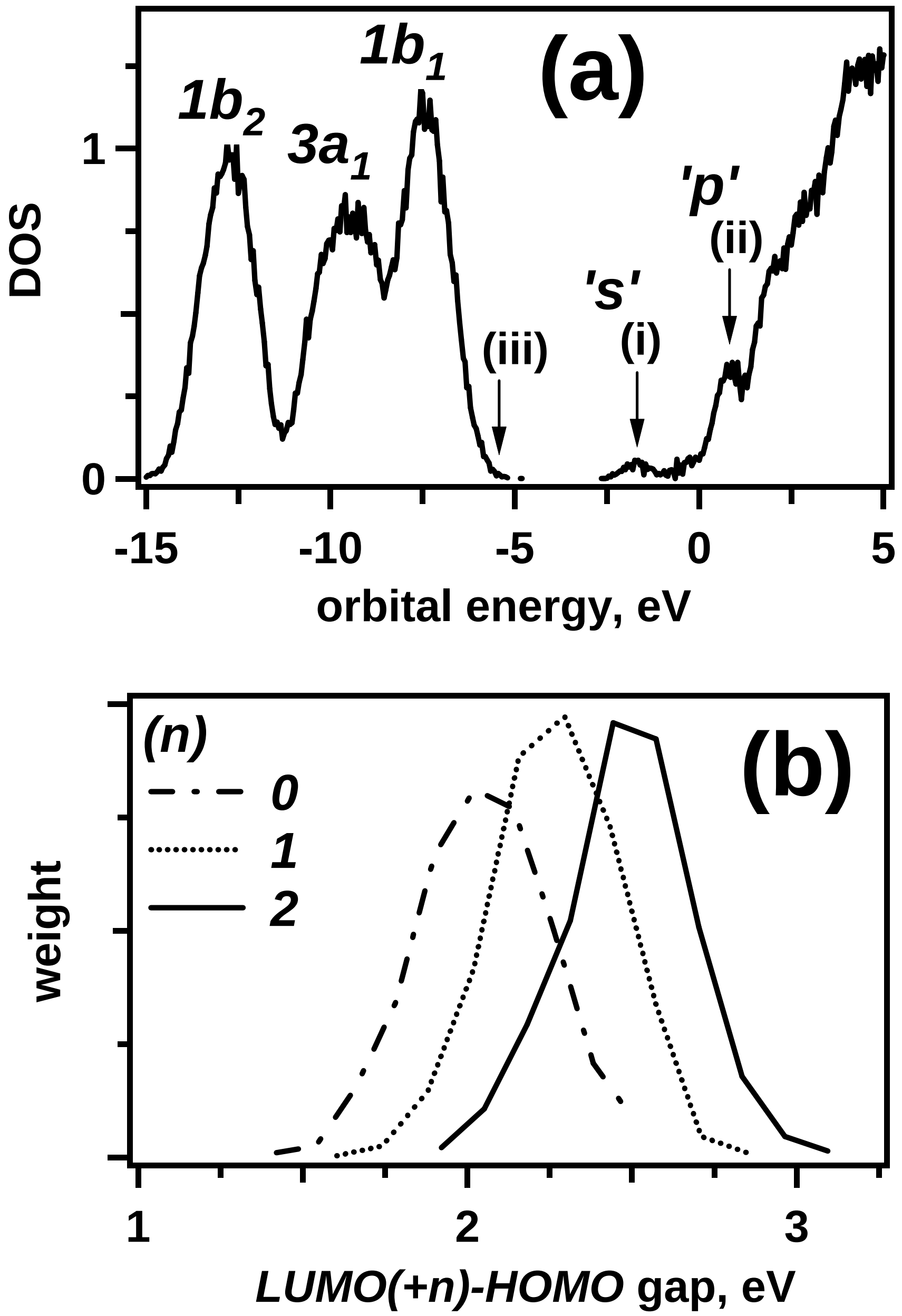

Figure 4. Shkrob

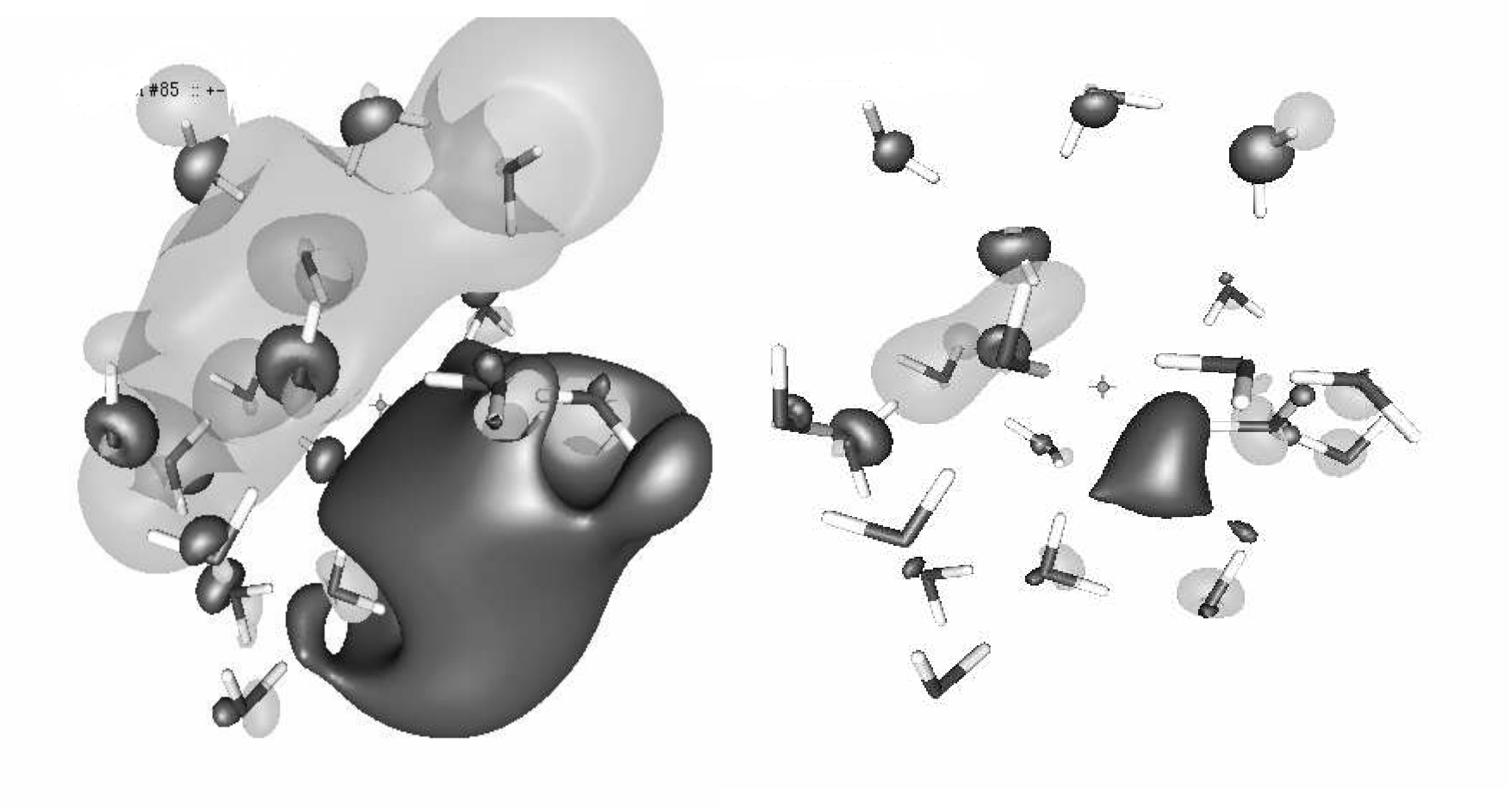

±0.01 ±0.03